\documentclass[twocolumn,prl,aps,nobalancelastpage,amsfonts,citeautoscript]{revtex4}

\usepackage{graphicx}
\usepackage{bm}
\usepackage{times}
\begin{document}

\title{Penetration depth study of superconducting gap structure of 2\textit{H}-NbSe$_2$}

\author{J.D. Fletcher,$^1$  A. Carrington,$^1$ P. Diener,$^2$ P. Rodi\`{e}re,$^2$\\ J.P.
Brison,$^2$  R. Prozorov,$^3$ T. Olheiser$^4$ and R.W. Giannetta$^4$} \affiliation{$^1$ H.H. Wills
Physics Laboratory, University of Bristol, Tyndall Avenue, BS8 1TL, United Kingdom.
\\$^2$CNRS-CRTBT, BP 166, 38042 Grenoble Cedex 9, France.
\\$^3$Department of Physics and Astronomy and Ames Laboratory, Iowa State University, Ames, Iowa
50011.
\\$^4$Department of Physics, University of Illinois at Urbana-Champaign, 1110 West Green St.,
Urbana, 61801 Illinois.}

\date{\today}

\begin{abstract}
We report measurements of the temperature dependence of both in-plane and out-of-plane penetration
depths ($\lambda_a$ and $\lambda_c$ respectively) in 2\textit{H}-NbSe$_2$. Measurements were made
with a radio-frequency tunnel diode oscillator circuit at temperatures down to 100~mK. Analysis of
the anisotropic superfluid density shows that a reduced energy gap is located on one or more of the
quasi-two-dimensional Nb Fermi surface sheets rather than on the Se sheet, in contrast to some
previous reports. This result suggests that the gap structure is not simply related to the weak
electron-phonon coupling on the Se sheet and is therefore important for microscopic models of
anisotropic superconductivity in this compound.
\end{abstract}
\maketitle

The transition metal dichalcogenide 2\textit{H}-NbSe$_2$ is unusual in that an incommensurate
charge density wave state ($T_{\rm CDW}<33~$K) coexists with superconductivity ($T_c\simeq 7.1~$K)
 \cite{MonctonAD77}. Recently, there has been renewed interest in this compound
because its unusual superconducting properties, such as the field dependence of the thermal
conductivity in the vortex state \cite{BoakninTPHRHSTSHB03}, bear a striking resemblance to the
two-gap superconductor MgB$_2$. Many experimental probes have found evidence for strong gap
anisotropy in 2\textit{H}-NbSe$_2$,  but it remains unclear whether this results from there being
different gaps on each Fermi surface sheet or some other form of gap anisotropy.  Determining the
nature of the gap anisotropy is an important step toward understanding its microscopic origin. Is
it related to the CDW, or does it rather result from an anisotropic electron-phonon interaction as
in MgB$_2$ \cite{LiuMK01}?

Angle resolved photoemission spectroscopy (ARPES) measurements \cite{YokoyaKCSNT01} performed at
$0.8~T_c$ have found that the superconducting gap is very small (below the experimental resolution
of 0.16 $k_{\mathrm B}T_c$) on the small, pancake-like Fermi surface sheet associated with the Se
$p$-orbitals. The gaps on the quasi-two-dimensional sheets, derived from the Nb $d$ orbitals, were
much larger ($1.6\pm0.16k_{\mathrm B}T_c$). The strength of the electron-phonon interaction
($\lambda_{ep}$) is also different between these sheets \cite{CorcoranMOPSTHGG94, VallaFJGMSAB04},
being rather weak on the Se sheet. This result suggests a similarity to MgB$_2$ where a smaller
superconducting gap on the $\pi$ sheets \cite{LiuMK01} compared to that on the $\sigma$ sheets
results from the being a large difference in $\lambda_{ep}$ between these sheets combined with very
weak $\sigma-\pi$ scattering.  The weak interband scattering is the crucial factor which produces
the possibly unique, two-gap superconducting state in MgB$_2$. It is as yet unclear whether a
similar situation exists in 2\textit{H}-NbSe$_2$.

We have performed measurements of the temperature dependence of the superfluid density to
investigate the gap structure in 2\textit{H}-NbSe$_2$. In an anisotropic superconductor, the
temperature dependent components of the superfluid density $\rho_i=1/\lambda_i^2(T)$ ($i=a,c$) are
sensitive to the distribution of the size of the superconducting gaps on the Fermi surface
\cite{ChandrasekharE93}. Here we present measurements of the temperature dependence of both the
in-plane and \emph{c}-axis magnetic penetration depths [$\lambda_a(T)$ and $\lambda_c(T)$] of
NbSe$_2$ from 100 mK up to $T_c$. We find that our results are not compatible with the
superconducting gap anisotropy as determined by the ARPES experiment \cite{YokoyaKCSNT01}. Our data
show that there is a reduced energy gap on one or more of the Nb Fermi surface sheets, but the gap
on the Se sheet is not smaller than this, and may even be larger.

In order to correctly interpret our measurements it is essential to have a good understanding of
the normal state electronic structure. Band structure calculations within the local density
approximation \cite{CorcoranMOPSTHGG94,JohannesMH06} show that three bands cross the Fermi level,
giving rise to five Fermi surface sheets.  Near the $\Gamma$ point there is a small pancake-like
sheet which derives mostly from the Se $p$ bands (band 16) \cite{JohannesMH06}. The other four
sheets derive from the Nb $d$ bands and are weakly warped tubes running along the \emph{c}-axis,
centered on the $\Gamma$ and $K$ points. The two surfaces derived from the bonding Nb $d$ band (17a
and 17b) are significantly more warped than those from the antibonding Nb $d$ band (18a and 18b).

This band structure has been found to be in good overall agreement with ARPES
\cite{VallaFJGMSAB04,YokoyaKCSNT01} and de Haas-van Alphen (dHvA) measurements
\cite{CorcoranMOPSTHGG94}. Significantly, the formation of the CDW state does not seem to lead to
major Fermi surface reconstruction.  dHvA measurements have only resolved signals originating from
the small Se pancake and show that, although the shape is similar to bandstructure calculations,
its size is somewhat smaller. The dHvA and ARPES results show that the mass renormalization factor
$(1+\lambda_{m^*})$ varies considerably between sheets.  It is largest on the Nb sheet 17b
($\lambda_{m^*}\simeq 1.9$ from ARPES \cite{VallaFJGMSAB04}) and smallest on the Se sheet 16
($\lambda_{m^*}\simeq 0.3$ from dHvA \cite{CorcoranMOPSTHGG94}). On the other Nb sheets
$\lambda_{m^*}\simeq 0.85$ \cite{VallaFJGMSAB04}.

In the clean local limit, $\rho_i(0)$ is related to the unrenormalized plasma frequencies
$\omega_{p,i}$ (in eV) by $\rho_s=\left(\omega_{p,i} e/c\hbar\right)^2(1+\lambda_{m^*})^{-1}$
\cite{ChandrasekharE93}. Band structure calculations of $\omega_{p,i}$ show that the Se sheet
contributes only $\sim 2$\% to $\rho_a(0)$, but $\sim$ 80\% to $\rho_c(0)$ (this reduces to
$\sim$50\% if the Se band is shifted slightly to agree with the dHvA results \cite{JohannesMH06}).
This shows that $\rho_a(T)$ is only sensitive to the gap on the quasi-2D Nb cyclinders, while
$\rho_c(T)$ is sensitive to the gap on both the Se pancake sheet and the Nb sheets in comparable
proportions.

The temperature dependence of the penetration depth was measured in single crystals of NbSe$_2$
using a radio frequency (RF) resonant LC circuit driven by a tunnel diode \cite{CarringtonGKG99},
operating at $\sim$12 MHz and low magnetic field ($H_{\rm RF}\simeq 10^{-2}$ Oe $\ll H_{c1}$). The
sample was attached with vacuum grease to a sapphire rod, the other end of which is attached to a
temperature controlled stage. The sample is placed in a copper solenoid which forms the inductor in
the LC circuit. Changes in the resonant frequency of the circuit ($\Delta F$) are directly
proportional to changes in the field penetration in the sample.

For $H\|$c the screening currents flow only in the basal plane and $\Delta F=\alpha\Delta
\lambda_{a}$, where the geometric factor $\alpha$ is estimated using the technique described in
Ref. \cite{ProzorovGCA00}. For $H\bot c$ currents flow both in the plane and along the
\emph{c}-axis, and (neglecting the small contribution from the top and bottom faces) for a
rectangular sample with dimensions $l_x$, $l_y$ and $l_z$ [$x$,$y$ are in-plane and $ l_x,l_y,l_z
\gg \lambda$] $\Delta F$ for $H\|y$ is given by \cite{ProzorovGCA00}:
\begin{equation}
\Delta F /\Delta F_0 = 2\Delta\lambda_a/l_z+2\Delta\lambda_c/l_x. \label{Eqcal}
\end{equation}
Here $\Delta F_0$ is the frequency shift obtained when the sample is completely removed from the
coil and accounts for the sample demagnetizing factor as well as the coil calibration factor.

Measurements were conducted on samples from three different sources (Tsukuba, Lausanne, Bell labs)
in three different laboratories. All samples were grown via the usual iodine vapor transport
technique and are known to be of high quality, having a high $T_c\simeq 7.1$K and low residual
resistances (RRR$\simeq$ 40-80). Some of the samples were from the same batch as those used for
dHvA measurements \cite{CorcoranMOPSTHGG94}. Experiments in Bristol were performed in either a
dilution fridge ($T_{\rm min} \simeq$ 100~mK) or a pumped $^4$He cryostat ($T_{\rm min}\simeq$
1.3~K). In Grenoble and Urbana-Champaign a $^3$He cryostat was used ($T_{\rm min}\simeq$ 0.5~K). In
total more than 15 samples were measured.

\begin{figure}
\center
\includegraphics*[width=6.5cm]{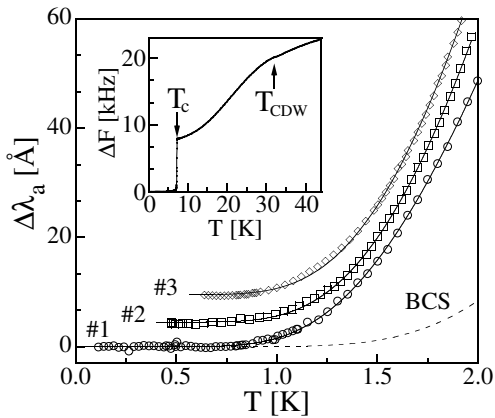}
\caption{Temperature dependence of the in-plane penetration depth, $\Delta\lambda_a$ in single
crystals of NbSe$_2$. Data for three different samples are shown, offset for clarity. The absolute
scale of $\Delta\lambda_a$ is most accurate for sample 1 (see text). For sample 2 the data have
been divided by 1.3.  Inset: Frequency shifts at higher temperature showing the charge density wave
transition.}\label{figinplane}
\end{figure}

Fig.\ \ref{figinplane} shows the low temperature variation of the in-plane penetration depth,
which, as mentioned above, probes mainly the excitations on the Nb sheets.  Data are shown for
three samples measured in different laboratories. Sample 1 (Bristol) was very thin ($0.48\times
0.55 \times 0.011~$mm$^3$) with $H\bot c$, hence the contribution from $\Delta \lambda_c$ is
negligible ($<2$\%). Samples 2 (Urbana/Ames) and 3 (Grenoble) were thicker, having $l_x\simeq l_y =
2.3l_z$, and were measured with $H\|c$. All three curves are fitted to the expression
\begin{equation}
\frac{\Delta\lambda}{\lambda(0)}\simeq\sqrt{\frac{\pi\Delta_0}{2T}}\exp
\left(-\frac{\Delta_0}{T}\right), \label{Eqbcs}
\end{equation}
which approximately reproduces the full solution to the BCS equations for $T\lesssim T_c/3$. For
all samples we find $\Delta_0=1.1 \pm 0.1$~$k_{\mathrm B}T_c$ (the quoted error includes a weak
dependence on fitting range). In a multigap superconductor, or one with a distribution of gap
values, at low temperature $\Delta_0$ is a measure of the smallest gap. For example, in MgB$_2$ a
low temperature fit measures the gap on the $\pi$ bands, which is much lower than the BCS weak
coupling value \cite{ManzanoCHLYT02}. Combined with our knowledge of the normal-state electronic
structure, the data clearly show the presence of excitations with an energy gap much smaller than
the weak coupling BCS value ($\Delta_0=1.76~k_{\mathrm B}T_c$) on the quasi-2D Nb sheets of Fermi
surface.

The temperature dependence of  $\Delta\lambda_a$ was very similar in all samples. We find that the
absolute values of $\Delta\lambda_a$ measured with $H\|c$ are sometimes higher than those for
$H\bot c$ (e.g., sample 2 in Fig.\ \ref{figinplane}).  This is in the opposite sense to that
expected from the additional $\lambda_c$ contribution for $H\bot c$. It is likely that this results
from the mica-like morphology of the crystals. Although very flat (001) faces may be prepared by
cleaving, cutting the crystal perpendicular to this direction produces splintered edges which may
have a larger effective area than their geometric cross-section (and hence larger effective field
penetration). For this reason, we believe the calibration factor is most accurately determined for
the \textit{thin} sample with $H\bot c$ (sample 1 in Fig.\ \ref{figinplane}) where $\sim$98\% of
the signal comes from the flat faces and demagnetization effects are small.

\begin{figure}
\center \includegraphics*[width=7cm]{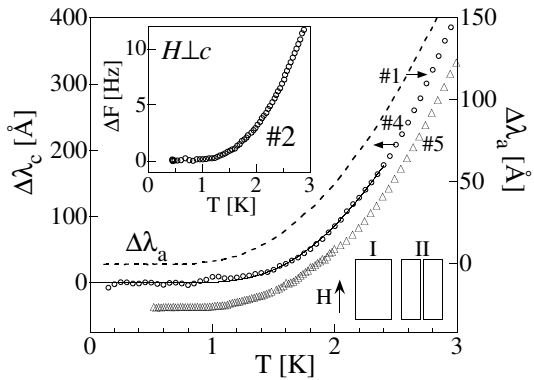} \caption{Low temperature behavior of $\Delta\lambda_c$
determined by cutting the sample (see text). Data for two different samples are shown (samples 4
and 5). For sample 5 the data have been divided by 2 and offset for clarity. The results for
$\Delta\lambda_a(T)$ (sample 1) are shown for comparison (dashed line). The solid line show a fit
to Eq.\ (\ref{Eqbcs}) up to 2.4~K. The lower inset shows the geometry of the samples before (I) and
after (II) cutting. Upper Inset: Raw measured frequency shift for a thick sample with $H\bot c$
(same as sample 2 in Fig.\ \ref{figinplane}).} \label{figlambdaC}
\end{figure}

We now turn to an analysis of $\lambda_c$ which probes the gaps on both the Se and Nb sheets. If
the gap on the Se sheet is significantly smaller than that on the Nb sheets \cite{YokoyaKCSNT01} we
would expect the temperature dependence of $\lambda_c$ to differ markedly from that of $\lambda_a$.

For \textit{thick} samples with $H\bot c$ a significant proportion of the total measured $\Delta
F(T)$ comes from currents running along the $c$-axis. In the inset to Fig.\ \ref{figlambdaC} we
show data for sample 2 where $l_z/l_x \simeq 0.44$. $\Delta F$ is temperature independent for
$T\lesssim 1.2$~K. This clearly shows that the gap is not substantially smaller on the Se sheet
than on the Nb sheets.

By measuring $\Delta F$ with H$\| x$ before and after cutting the sample in half along the field
direction (see Fig.\ \ref{figlambdaC}) we are able to make a more precise determination of
$\Delta\lambda_c(T)$. In principle, the in-plane contribution from the large faces is unchanged
after cutting, so the difference between frequency shifts is due to $c$-axis screening currents on
the newly exposed surfaces [Eq.\ (\ref{Eqcal})]. In practice, there may remain some contribution
from $\Delta\lambda_a$ because the effective edge area may change when the sample is cut due to
splintering. However, we found that when the sample was cut in two stages (first in half then in
quarters along the field direction), each cut resulted in an identical additional contribution to
$\Delta F(T)$.

In the main part of Fig.\ \ref{figlambdaC} we show our result for the low temperature behavior of
$\Delta\lambda_c$ determined by the cutting method in two samples \cite{dimensions}.
$\Delta\lambda_c$ has a very similar temperature dependence to $\Delta\lambda_a$.  There is a
variation in the absolute magnitude of $\Delta\lambda_c$ between samples, possibly due to a stray
contribution from $\lambda_a$, but the measured temperature dependence of $\lambda_c$ is very
similar.  In this temperature range $\Delta\lambda_c(T)$ displays exponential behavior at low
temperature indicating a minimum gap, $\Delta_0=1.2\pm0.2~T_c$, similar to that found in
$\Delta\lambda_a(T)$. The absence of any strong temperature dependence in $\Delta\lambda_c(T)$ at
low temperature and the similarity to $\Delta\lambda_a(T)$ demonstrates our key result, that
\textit{the gap on the Se sheet is at least as large as that on the Nb sheets}.

\begin{figure}
 \center \includegraphics*[width=6.4cm]{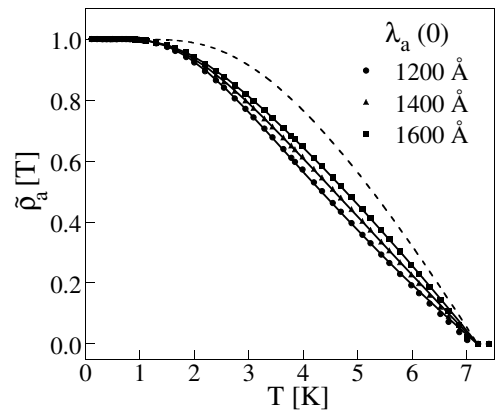}
\caption{Temperature dependence of the normalized in-plane superfluid density $\tilde{\rho}_a$
calculated using the $\lambda_a(0)$ values indicated.  The solid lines are fits to the two gap
model described in the text. The dashed line is the weak coupling BCS prediction with
$\Delta_0=1.76~k_{\mathrm B}T_c$.}\label{figrhoa}\end{figure}

In Fig.\ \ref{figrhoa} we show the normalized in-plane superfluid density, $\tilde{\rho_a} =
\lambda_a(0)^2/\lambda_a(T)^2$ for a range of values of $\lambda_a(0)$ close to that determined by
muon spin rotation measurements \cite{CallaghanLKS05} [$\lambda_a(0)=1250$~\AA]. As the data
clearly do not follow the standard BCS behavior we have attempted to model $\rho_a(T)$ using a
$\bm{k}$-dependent gap. In the clean local limit BCS theory $\rho_a$ is given by
\cite{ChandrasekharE93}
\begin{eqnarray}\label{eqrhoeq}
\rho_{a}(T)&=&\frac{\mu_0e^2}{4\pi^3\hbar}\left\{\oint dS_F \frac{v_{a}^2}{v}  \right.\nonumber\\
&&+ \left. 2\oint dS_F \frac{v_{a}^2}{v}\int_{\Delta_k}^\infty
\frac{df(\varepsilon)}{d\varepsilon}\frac{\varepsilon}{(\varepsilon^2-\Delta_k^2
)^{\frac{1}{2}}}d\varepsilon\right\}
\end{eqnarray}
where $dS_F$ is an element of Fermi surface, $v$ is the Fermi velocity, $f$ is the Fermi function
and $\Delta_k$ is the $\bm{k}$ dependent energy gap.

\begin{table}\begin{center}
\caption{Variation of the fit parameters to models of the in-plane superfluid density as a function
of the assumed zero temperature value of $\lambda_a$.  $\Delta_1$, $\Delta_2$, and $x$ are the
parameters of the two-gap model [Eq.\ (\ref{twogap})] and $\Delta(0)$ and $\epsilon$ are the
parameters of the 6-fold anisotropic gap model [Eq. (\ref{Eq6fold})]. The final line shows the
parameters found by fitting the same models \cite{xnote} to the heat capacity $C$ data of Ref.\
\onlinecite{SanchezJMBL95}.}\label{fitparams}

\begin{tabular}{cccccccc}
\hline\hline
&&\multicolumn{3}{c}{Two gap}&&\multicolumn{2}{c}{6-fold gap}\\
$\lambda_a(0)$&&$\frac{\Delta_1}{T_c}$&$\frac{\Delta_2}{\Delta_1}$&$x$&&$\frac{\Delta_{\rm
min}(0)}{T_c}$&$\frac{1+\epsilon}{1-\epsilon}$\\
1200\AA &&0.99&1.59&0.43&&0.94&1.74\\
1400\AA &&1.04&1.76&0.49&&0.91&2.07\\
1600\AA &&1.08&1.91&0.50&&0.91&2.33\\
\hline $C$&&1.31&1.76&0.30&&1.25&2.31
\\
\hline \hline\end{tabular}\end{center}

\end{table}

Our first approach is to assume that the gap on one, or more, of the two quasi-2D Nb sheets is
different to the others, but there is no variation of the gap within each sheet (similar to the
situation in MgB$_2$) \cite{BouquetWFHJJP01,ManzanoCHLYT02,FletcherCTKK05}.  Eq.\ (\ref{eqrhoeq})
then reduces to
\begin{equation}\label{twogap}
    \tilde{\rho}(T)=x\tilde{\rho}[T,\Delta_1(T)]+(1-x)\tilde{\rho}[T,\Delta_2(T)]
\end{equation}
where $x$ is the Fermi surface weight, which is proportional to the total superfluid density on
sheet(s) with the reduced gap, and $\Delta_{1,2}(T)=\Delta_{1,2}\Delta_{\rm BCS}(T)/\Delta_{\rm
BCS}(0)$, where $\Delta_{\rm BCS}(T)$ is the BCS weak coupling gap. The fits to $\rho_a(T)$ with
this model are excellent (see Fig. \ref{figrhoa}). The parameters are given in Table
\ref{fitparams}. The value of the small gap varies little with $\lambda_a(0)$ and is close to that
found from fitting Eq.\ (\ref{Eqbcs}) at low temperature. The gap ratio is
$\Delta_{2}/\Delta_{1}=1.8\pm0.2$. $x\simeq 0.5$ indicating that the sheets with the reduced gap
contribute $\sim 50$\% to the total in-plane superfluid density.

Our second approach considers a gap with 6-fold symmetry (NbSe$_2$ is hexagonal) on all of the Nb
sheets,
\begin{equation}\label{Eq6fold}
   \Delta(\phi,T)=\Delta_{\rm min}(T) \frac{[1 + \epsilon \cos(6\phi)]}{1-\epsilon}
\end{equation}
where, as above, $\Delta_{\rm min}(T)$ is assumed to have the BCS temperature dependence and $\phi$
is the in-plane angle (we approximate the Nb sheets as simple tubes). We find that this model fits
$\rho_a(T)$ as well as the two gap model above. The two free parameters [$\Delta_{\rm min}(0)$ and
$\epsilon$], which vary slightly with the assumed $\lambda_a(0)$, are given in Table
\ref{fitparams}. The ratio of the maximum to minimum gap is $(1+\epsilon)/(1-\epsilon)=2.0\pm0.3$,
which is close to that found for the 2 gap fits.

A two-gap analysis of $\rho_c(T)$ has also been performed. The minimum gap is found to be
$\Delta_1\simeq 1.2\pm0.2~k_BT_c$, in agreement with the low temperature fit of
$\Delta\lambda_c(T)$ to Eq.\ (\ref{Eqbcs}).  The value of the larger gap ranges from 1.2 to 1.8
$k_BT_c$ depending on our assumed value of $\lambda_c(0)$ and the relative contribution of the Nb
and Se sheets to $\rho_c(T)$.

By analyzing the zero field specific heat data of Ref.\ \onlinecite{SanchezJMBL95} with the same
models \cite{BouquetWFHJJP01} we find that the gap values are $\sim 30\%$ higher than those from
the superfluid density analysis. However, the values of the gap anisotropy are very similar (see
Table \ref{fitparams}) \cite{xnote}.

The above analysis is unable to differentiate between a discrete two-gap model and a continuously
varying gap. However, the influence of impurities strongly favors the two-gap model. We find no
difference in the behavior of $\lambda_a$ for samples with RRR in the range 40-80. A similar
conclusion was found from analysis of the field dependence of the specific heat
\cite{HanaguriKTNTK03}.  The criterion for observing an anisotropic (or sheet dependent) gap is
 $\hbar\tau^{-1}\ll \sqrt{\langle\Delta\rangle\delta\Delta}$, where $\tau^{-1}$
is the impurity scattering rate, $\langle\Delta\rangle$ is the average gap and $\delta\Delta$ is
its variation over the Fermi surface \cite{MazinAJGDK04}. Using the parameters of either our fits
yields $\sqrt{\langle\Delta\rangle\delta\Delta}\simeq 7$K. For a sample with RRR of 40, we estimate
the \textit{isotropic} mean free path is $\sim 270$~\AA, and $\hbar\tau^{-1}=\hbar
v_{a}/(k_{\mathrm B}\ell) \simeq$ 27 K which is $\sim$4 times larger than
$\sqrt{\langle\Delta\rangle\delta\Delta}$. The scattering must therefore be \textit{highly}
anisotropic. A likely scenario is that there is reduced scattering between the two Nb bands. This
would allow two different gaps to exist on these pairs of Fermi surface sheets (as for MgB$_2$
\cite{MazinAJDKGKV02}). The reduced energy gap on one of the Nb sheets could be related to the CDW
state formation.

In conclusion, our measurements of the anisotropic superfluid density in NbSe$_2$ show evidence for
anisotropic, probably multi-band superconductivity. However, this is not of the form suggested by
some previous reports, where the small Se sheet is the only sheet with a reduced gap, and is
therefore not simply due to the weak electron-phonon coupling on the Se sheet. We hope our results
will stimulate theoretical investigations of the microscopic origin of the gap anisotropy, which
may be related to the formation of the CDW state.

We thank I.\ Mazin for useful discussions, as well as H.\ Berger, Y.\ Onuki and P.\ Gammel for
providing the NbSe$_2$ samples.


\begin{thebibliography}{20}
\bibitem[{dim(1900)}]{dimensions}
\bibinfo{note}{Samples 4, and 5 had the following respective dimensions ($a
  \times b \times c$) 0.76$\times$0.61$\times$0.06~mm$^3$;
  0.47$\times$0.45$\times$0.077~mm$^3$}.
\bibitem[{xno(1900)}]{xnote}
\bibinfo{note}{For the $C$ analysis $x$ is proportional to the relative
  contribution of the density of states of each band to the total.}
\end{thebibliography}
\end{document}